\documentclass[11pt]{article}
\usepackage{moriond,epsfig}

\def\be{\begin{equation}}
\def\ee{\end{equation}}
\def\bea{\begin{eqnarray}}
\def\eea{\end{eqnarray}}

\begin{document}
\vspace*{4cm}
\title{COMPARISON OF LISA AND ATOM INTERFEROMETRY FOR\\
                   GRAVITATIONAL WAVE ASTRONOMY IN SPACE}

\author{PETER L. BENDER }

\address{JILA, University of Colorado and National Institute of\\
Standards and Technology, Boulder, CO}

\maketitle\abstracts{
 One of the atom interferometer gravitational wave missions
        proposed by Dimopoulos {\it et al.}\cite{dim08} in 2008 was called AGIS-Sat.\,\,2.
        It had a suggested gravitational wave sensitivity set by the
        atom state detection shot noise level that started at 1 mHz,
        was comparable to LISA sensitivity from 1 to about 20 mHz, and
        had better sensitivity from 20 to 500 mHz.  The separation
        between the spacecraft was 1,000 km, with atom interferometers
        200 m long and shades from sunlight used at each end.  A
        careful analysis of many error sources was included, but
        requirements on the time-stability of both the laser wavefront
        aberrations and the atom temperatures in the atom clouds were
        not investigated.  After including these considerations, the
        laser wavefront aberration stability requirement to meet the
        quoted sensitivity level is about $1 \times 10^{-8}$ wavelengths, and is
        far tighter than for LISA.  Also, the temperature
        fluctuations between atom clouds have to be less than 1 pK.
           An alternate atom interferometer GW mission in Earth orbit
        called AGIS-LEO with 30 km satellite separation has been
        suggested recently.  The reduction of wavefront aberration
        noise by sending the laser beam through a high-finesse mode-scrubbing optical cavity is discussed briefly, but the
        requirements on such a cavity are not given.  Unfortunately,
        such an Earth-orbiting mission seems to be considerably more
        difficult to design than a non-geocentric mission and does
        not appear to have comparably attractive scientific goals.}

\section{Introduction}

The purpose of this paper is to discuss some proposals that have been made to
use atom interferometry in space missions to observe gravitational waves.
Three specific space mission candidates were proposed by Dimopoulos {\it et al.}
in 2008.\cite{dim08}  The missions were called Atom Gravitational wave Interferometric
Sensor (AGIS), Satellite 1, 2, and 3 ({\it i.e.}, AGIS-Sat. 1, etc.).  It appears
useful to compare these missions with the Laser Interferometer Space Antenna
(LISA) gravitational wave mission\cite{danz03,bend10} that has been studied extensively as a
proposed joint mission of the European Space Agency and NASA.  The
AGIS-Sat. 2 mission has a nominal sensitivity curve closest to that of LISA,
and it will be the main mission discussed here.

   After reading ref. 1 and attempting to obtain more information about the
proposed missions, it became clear that there were quite severe additional
requirements needed in order to meet the given nominal sensitivities.  Thus a
Comment on the paper by Dimopoulos et al. was prepared and submitted to
Physical Review D in August, 2010.   A somewhat modified version of this
Comment\cite{bend11} has now been accepted for publication.

   In September, 2010, a paper by Hogan {\it et al.}\cite{hogan10} describing a proposed AGIS
mission in low Earth orbit called AGIS-LEO was placed on arXiv.  For this
mission, the optimum part of the nominal sensitivity curve is moved up in
frequency to 0.03 to 10 Hz, compared with 0.003 to 0.5 Hz for AGIS-Sat. 2,
and the sensitivity is about a factor 20 worse.  The main orbit geometry
considered was a leader-follower configuration on a circular orbit at nearly
constant altitude.

   The proposed AGIS-Sat. 2 mission and the requirements to meet its
sensitivity goals will be described in Section 2.  This will be followed by a
discussion of the proposed AGIS-LEO mission in Section 3.  Then, a brief
comparison with the requirements of LISA will be given in Section 4.

\section{AGIS-Sat. 2 Mission}
In ref. 1, each of the three space missions proposed was assumed to make use
of short sequences of laser pulses at three different times separated by
times $T$ to carry out the atom interferometry.  Satellites at each end of a
path of length $L$ would prepare atom clouds with temperatures of 100 pK and
send them out at a rate of one per second along the path.  Pulsed laser beams
from one end would provide the light pulses for the atom interferometry, and
a continuous laser beam from the other end would provide the phase reference
needed to permit correlation of the results obtained by the atom
interferometers at the two ends.

   The proposals for AGIS-Sat. 2 and AGIS-Sat. 3 assumed times $T$ between the
three pulse sequences of 100 s, atom interferometer path lengths of 100 to
200 m at each end, and a difference of 200 to 400 photon momenta between
momenta transferred to the two split parts of the atom wavefunctions by the
first of the short laser pulse sequences.  However, a factor 10 larger value
for the distance $L$ between satellites was assumed for AGIS-Sat. 3, and a
factor 10 better phase sensitivity for detecting differences in the atom
populations in two ground-state sublevels at the end of each atom
interferometer, leading to about a factor 50 better nominal gravitational
wave sensitivity than for AGIS-Sat. 2.

   In ref. 4, the proposal for AGIS-Sat. 3 was considered.  However, ref. 1
says that AGIS-Sat. 3 ``is an aggressive possibility that might be realizable
in the future."  Since the sensitivity for AGIS-Sat. 2 is comparable with
that for LISA from about 1 to 20 mHz, and since AGIS-Sat. 3 appears to be
much more difficult to implement, attention will be focused on the
AGIS-Sat. 2 proposal in this paper.  For AGIS-Sat. 1, the nominal
gravitational wave sensitivity is a factor of roughly 20 worse than for
AGIS-Sat. 2 down to about 0.03 Hz, and much worse at lower frequencies.
Whether there is a science justification for such a mission appears to be
uncertain.

   It is stated clearly in ref. 1 that the nominal gravitational wave
sensitivities given are only those due to the statistical uncertainties in
atom sublevel populations determined at the ends of the atom interferometers.
A large number of other error sources are considered, but none are estimated
to exceed the statistical uncertainties.  However, two additional error
sources that were not considered are the subject of ref. 4.  The first of
these is laser wavefront aberration variations over periods of 1 to 200 s.
For a number of error sources considered in ref. 1, there is a strong
cancellation of the errors because they are closely the same for the atoms in
the two atom interferometers.  For example, the effect of laser phase noise
at fairly low frequencies is reduced because the travel time between the two
interferometers separated by 1000 km is only 0.003 s.  However, this is not
true for laser wavefront aberrations.

   The expected size of the atom clouds is considerably less than the
suggested telescope diameter of roughly 1 m for AGIS-Sat. 2.  And there will
be a substantial reduction in the amplitude of the wavefront aberrations over
the 1000 km path length.  An estimate similar to that made in ref. 4 based on
primary spherical aberrations indicates that such aberration variations would
need to be kept down to 1x10-8 wavelengths in order to keep the
gravitational wave noise from this source down to that from the statistical
atom state sensing noise.

   The second additional error source is fluctuations from cloud to cloud in
the atom cloud temperatures.  For 0.001 wavelength of dc primary spherical
aberration in the initially transmitted laser beam, fluctuations of only
1 pK in the atom cloud temperature from cloud to cloud would substantially
increase the gravitational wave noise level.

\section{AGIS-LEO proposal}

The proposal in ref. 5 for a mission called AGIS-LEO was quite different.  To
reduce some of the effects of being in Earth orbit, the baseline length
between the satellites was reduced to 30 km and the time interval $T$ between
the different short sets of laser pulses applied to the atoms was reduced to
4 s.  In addition, the use of five short sets of pulses instead of three and
operation at 1,000 km altitude are assumed.  The disturbing effects are
mainly gradients in the Earth's gravity field and the Coriolis force.

   Although the suggested gravitational wave sensitivity for AGIS-LEO is
about a factor 20 worse than for AGIS-Sat. 2, the requirement on the laser
wavelength aberration fluctuations is slightly tighter because of the
satellite separation being only 30 km.  The use of a high-finesse
mode-scrubbing cavity is discussed, but no estimate of the possible level of
wavefront aberration noise from a suitable laser is given, and corresponding
requirements on the filter cavity performance are not considered.  The
conceptual design shown for a single AGIS-LEO telescope is a 30 cm diameter
off-axis Gregorian system, and 1 W of laser power is assumed.

   The possible use of a pinhole spatial filter at the real intermediate
focus of the telescope to eliminate wavefront errors from all optics and
lasers before the primary mirror is mentioned.  However, in view of the
suggested laser beam waist size of 10 cm and the 30 cm telescope diameter,
careful apodization of the beam from the telescope appears to be needed in
order to reduce the amplitude of near-field diffraction ripples, which would
affect the atom clouds in the near interferometer differently than those in
the far interferometer.

   For the laser wavefront aberration noise, some information is available on
the fluctuations in wavefront tilt\cite{kwee09} from a set of 8 lasers similar to those
that might be used in the Advanced LIGO program.  These lasers had roughly
2 W of output power, and similar ones may be used as the master lasers in the
laser amplifier or injection-lock configurations needed to get the required
high input power for Advanced LIGO.  The relative pointing fluctuations for
the lasers were measured at frequencies down to 1 Hz, and were much higher at
that frequency than at 3 Hz.

  In the AGIS-LEO proposal, a possible alternative interferometer laser beam
geometry is discussed.  In this approach, the atom optics laser beams can be
made to first propagate between two satellite stations along a path that is
displaced from the atoms before being redirected to interact with the atoms.
As a consequence, the first propagation segment would serve as a spatial
filter, allowing high frequency wavefront noise to diffract out of the beam.
It is suggested that ``If needed, this alternative beam geometry could be
used in conjunction with a mode-scrubbing cavity."  However, for the longer
wavelength wavefront aberrations such as variations in wavefront curvature,
it appears that a substantial reduction in aberration amplitude would also
lead to a significant reduction in the laser power.

   Because of the reduction in the time $T$ between short sequences of laser
pulses for AGIS-LEO, the tight requirement on the temperature differences
between the atom clouds in the two atom interferometers is removed.  However,
this requirement is replaced by a very tight requirement on the fluctuations
in mean radial velocity for the clouds of 10 nm/s.  This requirement comes
from item 12 in Table IV of ref. 5, and involves the Earth's gravity gradient
and the satellite orbital frequency, plus a factor $T^4$.  It is stated that
such requirements could be relaxed by a moderate reduction in $T$, but there
would be some reduction in the measurement bandwidth also.

   In Fig. 4 of ref. 5, signal strength curves are shown for four types of
gravitational wave sources.  One of these is white dwarf binaries at 10 kpc
distance.  However, such binaries would only be detectable by AGIS-LEO at
frequencies above about 0.03 Hz, and it is not clear that there are likely to
be any white dwarf binaries currently in the galaxy at frequencies higher
than this.  The other types of sources shown are inspirals of one solar mass
black holes into $10^3$ or $10^5$ solar mass black holes at distances of up to 10
Mpc, but the expected rates for such events is very low.  Thus it does not
appear that there is a substantial scientific case for such a mission based
on gravitational wave detection.

   A secondary objective for AGIS-LEO that is mentioned in ref. 5 is the
determination of time variations in the Earth's gravity field.  The GRACE
satellite mission currently is monitoring such variations, but is near the
end of its life.  The next mission after GRACE probably will still fly at
roughly 500 km altitude, but later missions with fairly simple drag-free
systems are expected to fly at about 300 km altitude.  This is because of the
importance of monitoring time variations in the higher harmonics of the
Earth's field, and thus of obtaining higher spatial resolution.  The 1,000 km
altitude for AGIS-LEO would be a substantial limitation, since for degree
100 harmonics the attenuation of the signal at that altitude would be a
factor 20,000 higher than at 300 km altitude.

\section{Comparison of the LISA and AGIS-Sat. 2 Missions}

A major difference between the LISA and AGIS-Sat. 2 missions is in the degree
of complexity.  For LISA, one of the two main mechanical requirements is to
be able to clamp the test masses during launch, and then release them
reliably later.  The other, because of LISA needing to have at least two
interferometer arms, is to be able to change the angle between the two
optical assemblies sending beams along the arms smoothly over about a degree
range during the year.  These are quite standard engineering design
requirements.  For laser interferometry, the requirement of about $2 \times 10^{-5}$
wavelength/$\sqrt{{\rm Hz}}$ accuracy in measuring distance changes down to about 1 mHz
does not come close to the state of the art at all, and the only challenge is
to accomplish this reliably over the whole mission lifetime with fairly
simple hardware.

   For AGIS-Sat. 2, even without the additional requirements discussed
earlier, there are many more and more challenging requirements.  For example,
$10^8$ atom clouds have to be prepared and cooled to 100 pK temperature at a
rate of one cloud per second.  The clouds then have to be moved 30 m or more
from the satellite, placed along the axis of the laser beams, and sent off
accurately along the desired path.  The velocities have to be different for
the different clouds in order to permit them to be interrogated separately.
And the population ratios of the atom ground-state sublevels have to be
determined to $1 \times 10^{-4}$ accuracy up to more than 100 m from the spacecraft.  No
sketch of what a satellite capable of accomplishing this might look like appears to have been
presented so far in descriptions of the proposed mission.

   There also appears to be a problem with the 200 atom clouds assumed to be
simultaneously in each interferometer.  If sequences of Bragg and/or Raman
pulses are used to apply 100 units of photon momentum to each part of the
atom wavefunction, with 1 W of laser power and 1 m diameter telescopes, and
the stimulated Rabi frequency is 100 Hz, the spontaneous emission rate for
the atoms appears to be too high.  The possibility of operating about 10
concurrent interferometers is stated in Section V A 3 of ref. 5, but it isn't
clear that 200 clouds can be handled simultaneously for the set of parameters
assumed for AGIS-Sat. 2, unless there has been an error in understanding the
calculations.

   For the additional requirement on reducing laser wavefront aberration
noise, it is not clear if the impact on the design of the satellites would be
substantial.  In principle, a fairly small filter cavity could do what is
needed if the aberration noise level of roughly 1 W lasers is low.  Other
aberrations besides wavefront tilt that may be important are variations in
wavefront curvature and beam center displacements.  The laser power would
only be a consideration if the finesse needed is fairly high.

   The wavefront aberration noise requirement for AGIS-Sat. 2 is much tighter
than for LISA because of the far shorter baseline between satellites.  For
the statistical limit on sensitivity from the atom sublevel measurements, the
very short de Broglie wavelength of the atoms is the relevant length scale.
However, when laser beams between spacecraft are used to provide the
reference for gravitational wave sensing, the laser wavelength becomes an
important scale for systematic measurement limitations.  Even for possible
LISA satellite separations as short as $1 \times 10^6$ km, the AGIS-Sat. 2 baseline is
a factor 1,000 shorter, and the sensitivity to wavefront aberration noise
would be increased by this factor

   For the requirement on the atom cloud temperature variations, it seems
difficult to see a solution other than reducing the time T substantially or
developing methods for extremely precise control of cloud temperatures.  In
Section IV B 5 of ref. 5, it is suggested that ``Spatially resolved detection
of the atom cloud can help mitigate the wavefront requirements that result
from spatially averaging."  However, even with an extra requirement for
measurement of the atom spatial distribution, this wouldn't help with
determining fluctuations in the atom cloud temperature, since such
measurements would be made only at the time of atom sublevel population
determination.

   In the Introduction to ref. 1, it is stated that the use of atom
interferometry ``leads to a natural reduction in many systematic backgrounds,
allowing such an experiment to reach sensitivities comparable to and perhaps
better than LISA's with reduced engineering requirements."  But, in fact,
nothing in that paper or in ref. 5 supports that claim.

\end{document}